# NEW OBSERVATIONS OF THE SOLAR 0.5–5 KEV SOFT X-RAY SPECTRUM

Amir Caspi[1][†], Thomas N. Woods[1], and Harry P. Warren[2]
[1]Laboratory for Atmospheric and Space Physics, University of Colorado, Boulder, CO 80303, USA
[†]now at Southwest Research Institute, Boulder, CO 80302, USA
[2]Space Science Division, Naval Research Laboratory, Washington, DC 20375, USA

## ABSTRACT

The solar corona is orders of magnitude hotter than the underlying photosphere, but how the corona attains such high temperatures is still not understood. Soft X-ray (SXR) emission provides important diagnostics for thermal processes in the high-temperature corona, and is also an important driver of ionospheric dynamics at Earth. There is a crucial observational gap between ~0.2 and ~4 keV, outside the ranges of existing spectrometers. We present observations from a new SXR spectrometer, the Amptek X123-SDD, which measured the spatially-integrated solar spectral irradiance from ~0.5 to ~5 keV, with ~0.15 keV FWHM resolution, during sounding rocket flights on 2012 June 23 and 2013 October 21. These measurements show that the highly variable SXR emission is orders of magnitude greater than that during the deep minimum of 2009, even with only weak activity. The observed spectra show significant high-temperature (5–10 MK) emission and are well fit by simple power-law temperature distributions with indices of ~6, close to the predictions of nanoflare models of coronal heating. Observations during the more active 2013 flight indicate an enrichment of low first-ionization potential (FIP) elements of only ~1.6, below the usually-observed value of ~4, suggesting that abundance variations may be related to coronal heating processes. The XUV Photometer System Level 4 data product, a spectral irradiance model derived from integrated broadband measurements, significantly overestimates the spectra from both flights, suggesting a need for revision of its non-flare reference spectra, with important implications for studies of Earth ionospheric dynamics driven by solar SXRs.

## 1. INTRODUCTION

The solar corona, at quiescent (non-flaring) temperatures of ~1–2 megaKelvin (MK), is ≳100× hotter than the underlying chromosphere and photosphere. This "coronal heating problem" remains one of the fundamental unanswered questions in solar (and stellar) physics (see, e.g., the review by Klimchuk 2006). Magnetohydrodynamic simulations and observations of convective flows (e.g., Welsch 2014, and references therein) suggest that the Sun's complex magnetic field is an efficient conduit for energy transport from the solar interior and subsequent storage in the corona. Models based on the impulsive dissipation of magnetic complexity through magnetic reconnection ("nanoflares," e.g., Parker 1988) suggest that coronal plasma should be routinely heated to flare-like temperatures, but with relatively low density (e.g., Cargill & Klimchuk 2004; Cargill 2014). In contrast, models based on the dissipation of Alfvén waves predict relatively narrow distributions of coronal temperatures (e.g., Asgari-Targhi et al. 2013). However, the difficulty of measuring weak, high temperature emission has led to inconsistent results (see, e.g., Reale et al. 2009; Schmelz et al. 2009; Warren et al. 2012; Del Zanna & Mason 2014).

Soft X-ray (SXR; ~0.1–10 keV, or ~0.1–10 nm) emission is particularly sensitive to high-temperature plasma and is thus an important diagnostic of the coronal temperature distribution and elemental abundances, and their evolution. Spectrally-resolved observations are crucial for constraining the various coronal heating models (Winebarger et al. 2012). The *Reuven Ramaty High Energy Solar Spectroscopic Imager* (*RHESSI*; Lin et al. 2002) is highly sensitive to flare X-rays, even from microflares (e.g., Hannah et al. 2008), but is only marginally sensitive to quiescent SXR emission (McTiernan 2009). The extreme ultraviolet (EUV) spectrum is routinely measured by the EUV Variability Experiment (EVE; Woods et al. 2012) onboard the *Solar Dynamics Observatory* (*SDO*; Pesnell et al. 2012), but there are very few coronal EUV lines in the 5–10 MK range essential for probing active region heating.

A critical observational gap exists from ~0.2 to ~4 keV (~0.3–6 nm), between the usable ranges of EVE and *RHESSI*. Decades of spectrally-integrated broadband measurements cover this range, notably from the *Geostationary Operational Environmental Satellite* (*GOES*) X-ray Sensor (XRS; Garcia 1994), the XUV Photometer System (XPS; Woods et al. 2008) on multiple spacecraft, and the EUV SpectroPhotometer (Didkovsky et al. 2012) within *SDO*/EVE. However, these integrated observations provide little spectral information and often disagree with one another. Many fewer spectrally-resolved observations exist, including from the Bragg Crystal Spectrometer on *Yohkoh* (Culhane et al. 1991), the Solar Array for X-rays (Schlemm et al. 2007) onboard *MESSENGER*, and the Solar Photometer in X-rays (SphinX; Gburek et al. 2011) onboard *CORONAS-Photon*. All of these instruments had either limited spectral coverage or resolution, and none covered the full ~0.2–4 keV range; consequently, the spectral distribution in this range is still poorly quantified.

This uncertainty has important geospace implications, as photons at these energies are preferentially absorbed in the D- and E-regions of Earth's ionosphere. The resultant dynamics depend critically on the altitude where the SXR energy is absorbed, which, because of the steep photoionization cross-sections of atmospheric constituents, is determined largely by the (unknown) spectral distribution. This is especially important during solar flares, where the spectral variability is expected to peak around ~0.6 keV (Rodgers et al. 2006). Spectral irradiance models developed from integrated broadband measurements (e.g., XPS Level 4 — Woods et al. 2008; or the Flare Irradiance Spectral Model — Chamberlin et al. 2008) can disagree with one another by up to an order of magnitude, largely because they must assume a spectral distribution *a priori*. It has therefore proven difficult to reconcile observed ionospheric dynamics with those predicted using such models (e.g., Sojka et al. 2013).

The Amptek X123-SDD, a new SXR spectrometer providing the highest resolution and lowest energy threshold to date from a spectrally-resolved broadband instrument, can address these issues. We present spatially-integrated X123 observations from two sounding rocket flights, in 2012 and 2013, with differing solar activity. Thermal model fits to the data suggest non-negligible high-temperature (~5–10 MK) emission, and, for one flight, likely deviations from coronal abundances. The corresponding XPS L4 model spectra significantly exceed our observations, suggesting a need for adjustment to the model.

## 2. INSTRUMENT DETAILS

The Amptek X123-SDD[1] package includes a silicon drift detector (SDD) and two-stage thermoelectric cooler in a vacuum housing with a Be entrance window, a high-voltage power supply, and a full-featured multi-channel analyzer with many user-configurable options. Cooling the detector to ~ –50°C, and a smaller capacitance, together enable an improved ~0.15 keV FWHM resolution compared to the Si p-i-n technology flown previously (e.g., on SphinX). An 8–13 µm Be thickness and 500 µm Si depletion depth provide sensitivity to X-rays from ~0.5 keV up to ≳30 keV. The signal processing chain is fully digital except for the pre-amplifier, allowing faster photon counting than traditional analog electronics. Parallel fast and slow pulse shapers enable on-board pulse pile-up rejection, similar to *RHESSI* (Smith *et al.* 2002).

We operated X123 with 1024 channels covering ~0.5–30 keV, oversampling the resolution by ~5, with 1 s cadence (see §3 for a caveat regarding the first rocket flight). The detector gain and offset were calibrated with $^{55}$Fe and $^{241}$Am radioactive sources, yielding ~0.0296 keV/channel and ~ –0.11 keV, respectively. A circular, ~340 µm-diameter precision aperture ensured moderate count rates, preventing high detector deadtime and pile-up; the ~120 µm-thick tungsten aperture plate provided ≪1% stray light transmission for X-rays ≲30 keV. A stainless steel baffle restricted the field of view to ~ ±5°.

The end-to-end detector response (counts per photon, or quantum throughput) was calibrated using beamline 2 of the Synchrotron Ultraviolet Radiation Facility (SURF; Arp *et al.* 2011) at the National Institute of Standards and Technology. The SURF spectrum above ~0.5 keV is calibrated to ≲10% (M. Furst 2014, private communication). Figure 1 (left) shows an example count rate spectrum observed from the 408 MeV SURF beam, with a total count rate of ~2300 cts s$^{-1}$. The spectrum is normalized by synchrotron beam current as a proxy for beam intensity. Comparisons of spectra with varying count rates showed negligible pulse pile-up below ~$10^4$ cts s$^{-1}$.

Ideally, the instrument response could be determined by directly dividing the observed counts by the known input photon flux, but, in practice, this simple inversion is complicated by non-photopeak response elements (e.g., instrument resolution) and is noisy above ~2–3 keV due to counting statistics. Instead, we model the response using the Henke atomic scattering factors for Be and Si (Henke *et al.* 1993). We optimize the model parameters by convolving the known input spectrum with the modeled response to generate a predicted spectrum that is then fit against the observations. For simplicity, we assume a quasi-diagonal (photopeak-only, plus resolution broadening) response matrix. An effective Be filter thickness of ~15 µm was required to match the total count rate; the discrepancy from the Amptek-reported thickness of ~7–10 µm is consistent with known uncertainties in the Henke Be scattering factors and with manufacturing tolerances. The model also required a ~0.2 keV additional scalar offset in the energy-to-channel conversion, possibly due to electrical grounding issues in the SURF test setup (this offset did not appear in the lab or during flight, both using different support electronics). The Be thickness and energy offset were the only free parameters in the model.

An excess below the count-rate peak, from ~0.5 to ~1.0 keV, could not be explained by any Henke-derived model with physically-viable parameters, and is most likely from off-diagonal (energy loss) processes that begin to dominate as the diagonal (photopeak) response falls off at low energies (e.g., photoelectrons emitted from the Be filter interacting in the detector, or causing secondary bremsstrahlung from the aluminum vacuum housing; or escape of L-shell fluorescence photons excited within the Si detector; cf. Caspi 2010 for *RHESSI*). To approximately account for this, we derived the full instrument response in two pieces. Above ~1.1 keV, where the response is photopeak-dominated, we optimized the Henke-derived model as described above. Below this energy, we approximated a diagonal response through direct inversion, by deconvolving the instrument resolution from the observed counts and dividing by the known incident photon flux. Figure 1 (right) shows the Henke-only and "hybrid" responses determined this way, not including resolution broadening. By design, the hybrid-model-predicted spectrum matches the SURF observations well at all energies (Figure 1, left).

Although our directly-inverted response below ~1.1 keV is not fully physical, it is, by its nature, an upper limit for the true response. Conversely, the Henke-derived model is a lower limit, and hence the two models fully bound the true response to first order. A more precise determination would require physically modeling the instrument response, e.g., with GEANT (Agostinelli *et al.* 2003). Above ~1.1 keV, our calibration uncertainty is ≲10%, dominated by the uncertainty of the SURF spectrum.

## 3. OBSERVATIONS AND ANALYSIS

We flew X123 on two *SDO*/EVE sounding rocket calibration underflights: NASA 36.286 (2012 June 23, ~19:30 UT) and 36.290 (2013 October 21, ~18:00 UT), hereafter R20120623 and R20131021, respectively. Each flight provided ~5 minutes of accurately pointed solar observations. Figure 2 shows the F10.7 radio flux and the XPS ~0.1–7 nm broadband integrated SXR flux, along with ~0.1–7 nm broadband SXR images from EVE's Solar Aspect Monitor (Hock *et al.* 2012; Woods *et al.* 2012), for the two flights. R20120623 occurred during the minimum of a 27-day rotation period, with negligible on-disk activity and only weak emission from active regions on the limb, while R20131021 occurred during rising activity, with two large, strong on-disk active regions and additional emission from limb regions.

Figure 3 shows the spatially-integrated solar spectral irradiance (as photon flux) derived by dividing the observed count rate spectra from the two flights by the best-fit instrument response (Figure 1, right), then normalizing by aperture area and spectral bin width. The ~0.15 keV FWHM instrument resolution has *not* been deconvolved to avoid introducing additional noise. Limitations in electronics and telemetry during R20120623 required downsampling of the spectral and temporal binning to ~0.12 keV/channel (256 bins) and ~2 s, respectively. Both spectra average ~180 s of integrations centered on flight apogee (~280 km). Error bars are propagated from counting statistics only; below ~1 keV, the uncertainties are more accurately considered to be the difference between the models derived from the "hybrid" and Henke-only responses (solid and dashed curves, respectively). The total average count rates were ~1100 and ~3700 cts s$^{-1}$, respectively, so pulse

---
[1] http://www.amptek.com/products/x-123sdd-complete-x-ray-spectrometer-with-silicon-drift-detector-sdd/

pile-up is expected to be negligible.

Unsurprisingly, the R20131021 spectrum exceeds the R20120623 observations everywhere, by ~4×. The discrepancy is larger at higher energies, indicating a slightly higher average relative temperature for R20131021 (see below). The integrated 0.1–7 nm irradiances (Figure 2) differ by only ~2×, highlighting the greater variability at these higher energies (shorter wavelengths). For both flights, there is distinct emission up to ~5 keV. For comparison, Figure 3 also shows the ~1.2–3.0 keV spectral irradiance derived from SphinX observations of the 2009 deep minimum (Sylwester *et al.* 2012) and the upper limit of 3–6 keV quiet Sun flux derived from 2005–2009 *RHESSI* observations (Hannah *et al.* 2010). Both X123 spectra are significantly ($\gtrsim 10^{2-4}$×) brighter and harder than these truly-quiet Sun measurements, showing that SXR emission is strongly dependent on even "weak" activity.

To investigate the coronal temperature distribution, we forward-fit a photon spectral model, convolved with the instrument response, to the measured spectra. We fit both a simple two-temperature model (cf. Caspi & Lin 2010) and a power-law differential emission measure (DEM) model (cf. Cargill & Klimchuk 2004), including both continuum and line emission using the CHIANTI atomic database (v7.1; Dere *et al.* 1997; Landi *et al.* 2013) with ionization fractions from Mazzotta *et al.* (1998) and the standard coronal abundances (Feldman *et al.* 1992); importantly, a single-temperature model could not explain the data. For simplicity, we used IDL's built-in CURVEFIT function, fitting over only the ~1.1–4.0 keV range, where the instrument response is photopeak-dominated and counting statistics are sufficient.

Figure 4 shows the best-fit two-temperature and DEM models. Both spectra show dominant ~3 MK emission as would be expected above quiescent active regions. R20131021 shows significantly stronger high-temperature emission in both models compared to R20120623, consistent with our qualitative assessment from Figure 3. For both flights, the DEM fit is marginally better at higher energies ($\gtrsim 3.5$ keV). We feel the relatively high $\chi^2$ values are acceptable for this analysis given the simplistic models fit over a large number of data points, and the approximated response with unknown systematic uncertainties.

Spectral lines from hot ions are prominent in R20131021, including Mg XI (~1.35 keV), Si XII–XIII (~1.85 and ~2.2 keV), and S XIV–XVI and Ar XVI–XVIII (~2.4 and ~2.8–3.3 keV). Two lines appear to have higher-energy tails consistent with possible emission from Mg XII (~1.5 keV) and Si XIV (~2.0 keV), which may indicate that the Mg and Si ionization fractions need adjustment; this discrepancy persists even using the latest CHIANTI-default fractions. *RHESSI* hard X-ray observations show that R20131021 occurred during the decay phase of a microflare, but the *GOES* XRS 1–8 Å lightcurve indicates that the microflare contributes at most a few percent to the X123 spectrum. Examination of the time-resolved X123 spectra shows no significant evolution of the lines or continuum during the entire ~5-minute observation, suggesting that ionization non-equilibrium is not a likely contributor to this discrepancy.

Importantly, while the default coronal abundances yield an acceptable fit to R20120623, they do not do so for R20131021, where the data require that the abundance of elements with a low first-ionization potential (FIP; e.g., Laming 2004) be *reduced* to ~0.4× the default coronal values, particularly to match the Si XIII line at ~1.85 keV. Although correlated errors between the fitting parameters and CURVEFIT's simplistic sampling of $\chi^2$ space make it difficult to obtain rigorous error bars, from the minimum envelope of the reduced $\chi^2$ we conservatively estimate uncertainties of ±0.05 in the low-FIP scale factor, yielding best-fit values of near-coronal ~0.85–0.95 for R20120623, and significantly below-coronal ~0.35–0.45 for R20131021. The same scalar factor was applied uniformly to the prominent low-FIP elements Fe, Ni, Mg, Si, and Ca; the "mid-FIP" S was adjusted by half the factor. High-FIP elements, including C, Ne, Ar, and Co, were not adjusted.

## 4. SUMMARY AND CONCLUSIONS

The Amptek X123-SDD offers an essential advancement in measuring the solar SXR spectrum in the poorly-observed range of ~0.2–4 keV. Our X123 observations provide the highest resolution and lowest energy threshold to date from any broadband SXR spectrometer, and show that, even for weak activity, the SXR irradiance is orders of magnitude higher than during "spotless" periods such as the deep minimum of 2009.

The observed spectra are well fit by either a two-temperature or a power-law DEM model. Significantly, both models indicate the presence of high-temperature (5–10 MK) plasma, for two disparate activity levels. The emission measure at these temperatures is orders of magnitude smaller than at 2–3 MK, but is potentially consistent with impulsive heating models. For both power-law DEMs, we obtain indices of ~6, roughly consistent with the slope of 11/2 predicted by Cargill & Klimchuk (2004). However, the relationship between the DEM derived from disk-integrated observations and the theoretical distribution derived for an individual loop is not straightforward, and needs to be considered in more detail. Importantly, SphinX observations of the quiet Sun show negligible evidence of high-temperature emission (Sylwester *et al.* 2012), suggesting that different heating processes may dominate in the quiet network versus active regions, although this may be limited by SphinX's sensitivity and emphasizes the need for future observations with much greater collecting area (e.g., by *NuSTAR*; Hannah *et al.* 2014).

Our observations illustrate the diagnostic power of this spectral range for studies of elemental abundances. While R20120623 is consistent with a coronal composition, R20131021 suggests reduced low-FIP abundances of only ~0.4× the typical value, corresponding to an enrichment (relative to the photosphere) of ~1.6. Prior studies of solar and stellar abundances have yielded (nominal) low-FIP enrichments of ~3-4 above quiescent active regions (e.g., Warren *et al.* 2012; Del Zanna & Mason 2014) and similar values for disk-integrated observations during moderate to high activity (e.g., Laming *et al.* 1995; Laming & Drake 1999), while intermediate enrichments of ~2 (e.g., Fludra & Schmelz 1999) or photospheric compositions (e.g., Warren 2014) have been observed during flares. Because the microflare during R20131021 contributes negligibly to the X123 spectrum and its evolution, our observed abundance variation is likely intrinsic to active region heating and is not a transient effect. The differences between R20120623 and R20131021 could therefore suggest a connection between coronal heating processes and composition. Nonetheless, additional data are required to investigate these relationships further. Future studies could also consider abundance variations for individual elements — the Mg and Si lines may be sufficiently unblended for such analysis.

Admittedly, our two-temperature and DEM models are almost certainly cruder approximations of a probably more complicated temperature distribution. Additionally, the data currently cannot rule out an additional non-thermal power-law component, which, if included, could potentially affect our fit abundance values. Nonetheless, this component would require a quite soft spectral index of ~7, and non-thermal emission has never before been observed from the quiescent Sun — such an observation would significantly constrain applicable coronal heating models. Simultaneous measurements of emission from the same thermal processes at different wavelengths, such as from EVE, would help to investigate this. A more rigorous analysis, adapting the multi-instrument DEM technique of Caspi *et al.* (2014) for X123 data, and including abundance fitting from Warren (2014), will be discussed in a future paper.

Our measurements highlight the need for improved spectral modeling in this energy range. Figure 4 shows that, while the total integrated ~0.1–0.8 nm (~1.5–12.5 keV) irradiance reported by *GOES* XRS (including a 30% correction factor; R. Viereck 2014, private communication) agrees with that derived from X123 to <10%, the XPS L4 model spectra substantially overestimate the measurements in both cases, by ~7.7× and ~3.8×, respectively. This disagreement has significant implications for products and analyses derived from XPS L4, e.g., studies of the ionospheric response to solar SXR loading. XPS L4 is derived empirically from combinations of pre-determined "reference" spectra — including CHIANTI quiet Sun and active region DEMs, and isothermal spectra with temperatures determined from *GOES* XRS (e.g., White *et al.* 2005) — added and scaled to match the XPS-observed integrated ~0.1–7 nm broadband irradiance. The XPS measurement uncertainty is only ~30% (Woods *et al.* 2008), thus the sizable model overestimate suggests that lower-temperature non-flare reference spectra are required to obtain agreement with the X123 data.

Our observations are limited to only two brief epochs, both quiescent. To more comprehensively address the open questions of coronal heating and solar-driven ionospheric dynamics requires significantly longer-term observations, including during flares. To that end, our X123 is being integrated into the *Miniature X-ray Solar Spectrometer* (*MinXSS*), a NASA-funded 3U CubeSat scheduled to launch from the International Space Station in mid-2015. Over its expected 6–12 month mission lifetime, *MinXSS* will greatly expand our measurements of this poorly-observed 0.5–5 keV energy range, and help to improve our understanding of both heating of the solar corona and the subsequent ionospheric response to its highly-variable SXR emission.

This work was supported by NASA contract NAS5-02140. We thank J. Stone for his analysis during the summer 2012 LASP REU program, and A. Y. Shih for many helpful discussions.

## REFERENCES

Agostinelli, S., Allison, J., Amako, K., *et al.* 2003, *NIMPA*, **506**, 250
Arp, U., Clark, C., Deng, L., *et al.* 2011, *NIMPA*, **649**, 12
Asgari-Targhi, M., van Ballegooijen, A. A., Cranmer, S. R., & DeLuca, E. E. 2013, *ApJ*, **773**, 111
Cargill, P. J., & Klimchuk, J. A. 2004, *ApJ*, **605**, 911
Cargill, P. J., 2014, *ApJ*, **784**, 49
Caspi, A. 2010, *PhD thesis*, Univ. California, Berkeley (arXiv: 1105.1889)
Caspi, A., & Lin, R. P. 2010, *ApJL,* **725**, L161
Caspi, A., McTiernan, J. M., & Warren, H. P. 2014, *ApJL*, **788**, L31
Chamberlin, P. C., Woods, T. N., & Eparvier, F. G. 2008, *SpWea*, **6**, 5001
Culhane, J. L., Hiei, E., Doschek, G. A., *et al.* 1991, *SoPh*, **136**, 89
Del Zanna, G., & Mason, H. E. 2014, *A&A*, **565**, A14
Dere, K. P., Landi, E., Mason, H. E., Monsignori Fossi, B. C., & Young, P. R. 1997, *A&A*, **125**, 149
Didkovsky, L., Judge, D., Wieman, S., Woods, T., & Jones, A. 2012, *SoPh*, **275**, 179
Feldman, U., Mandelbaum, P., Seely, J. F., Doschek, G. A., & Gursky, H. 1992, *ApJS*, **81**, 387
Fludra, A., & Schmelz, J. T. 1999, *A&A*, **348**, 286
Garcia, H. A. 1994, *SoPh*, **154**, 275
Gburek, S., Sylwester, J., Kowalinski, M., *et al.* 2011, *SoSyR*, **45**, 189
Hannah, I. G., Christe, S., Krucker, S., *et al.* 2008, *ApJ*, **677**, 704
Hannah, I. G., Hudson, H. S., Hurford, G. J., & Lin, R. P. 2010, *ApJ*, **724**, 487
Hannah, I., Marsh, A., Glesener, L., *et al.* 2014, *AGU Fall Meeting*, abstract #SH12A-04
Henke, B. L., Gullikson, E. M., & Davis, J. C. 1993, *ADNDT*, **54**, 181
Hock, R. A., Chamberlin, P. C., Woods, T. N., *et al.* 2012, *SoPh*, **275**, 145
Klimchuk, J. A. 2006, *SoPh*, **234**, 41
Laming, J. M. 2004, *ApJ*, **614**, 1063
Laming, J. M., & Drake, J. J. 1999, *ApJ*, **516**, 324
Laming, J. M., Drake, J. J., & Widing, K. G. 1995, *ApJ*, **443**, 416
Landi, E., Young, P. R., Dere, K. P., Del Zanna, G., & Mason, H. E. 2013, *ApJ*, **763**, 86
Lin, R. P., Dennis, B. R., Hurford, G. J., *et al.* 2002, *SoPh,* **210**, 3
Mazzotta, P., Mazzitelli, G., Colafrancesco, S., & Vittorio, N. 1998, *A&AS*, **133**, 403
McTiernan, J. M. 2009, *ApJ*, **697**, 94
Parker, E. N. 1988, *ApJ*, **330**, 474
Pesnell, W. D., Thompson, B. J., & Chamberlin, P. C. 2012, *SoPh*, **275**, 3
Reale, F., Testa, P., Klimchuk, J. A., & Parenti, S. 2009, *ApJ*, **698**, 756
Rodgers, E. M., Bailey, S. M., Warren, H. P., Woods, T. N., & Eparvier, F. G. 2006, *JGRA*, **111**, A10S13
Schlemm, C. E., Starr, R. D., Ho, G. C., *et al.* 2007, *SSRv*, **131**, 393
Schmelz, J. T., Saar, S. H., DeLuca, E. E., *et al.* 2009, *ApJL*, **693**, L131
Sojka, J. J., Jensen, J., David, M., *et al.* 2013, *JGRA*, **118**, 5379
Smith, D. M., Lin, R. P., Turin, P., *et al.* 2002, *SoPh*, **210**, 33
Sylwester, J., Kowalinski, M., Gburek, S., *et al.* 2012, *ApJ*, **751**, 111
Warren, H. P. 2014, *ApJL*, **786**, L2
Warren, H. P., Winebarger, A. R., & Brooks, D. H. 2012, *ApJ*, **759**, 141
Welsch, B. T. 2014, *PASJ*, in press (arXiv: 1402.4794)
White, S. M., Thomas, R. J., & Schwartz, R. A. 2005, *SoPh*, **227**, 231
Winebarger, A. R., Warren, H. P., Schmelz, J. T., *et al.* 2012, *ApJL*, **746**, L17
Woods, T. N., Chamberlin, P. C., Peterson, W. K., *et al.* 2008, *SoPh*, **250**, 235
Woods, T. N., Eparvier, F. G., Hock, R., *et al.* 2012, *SoPh*, **275**, 115

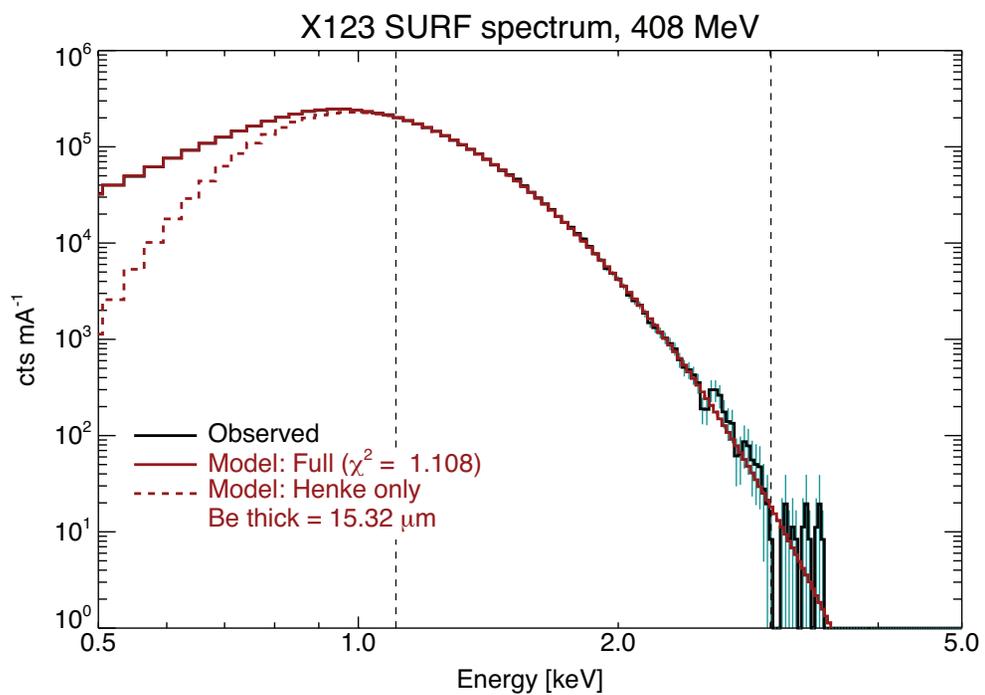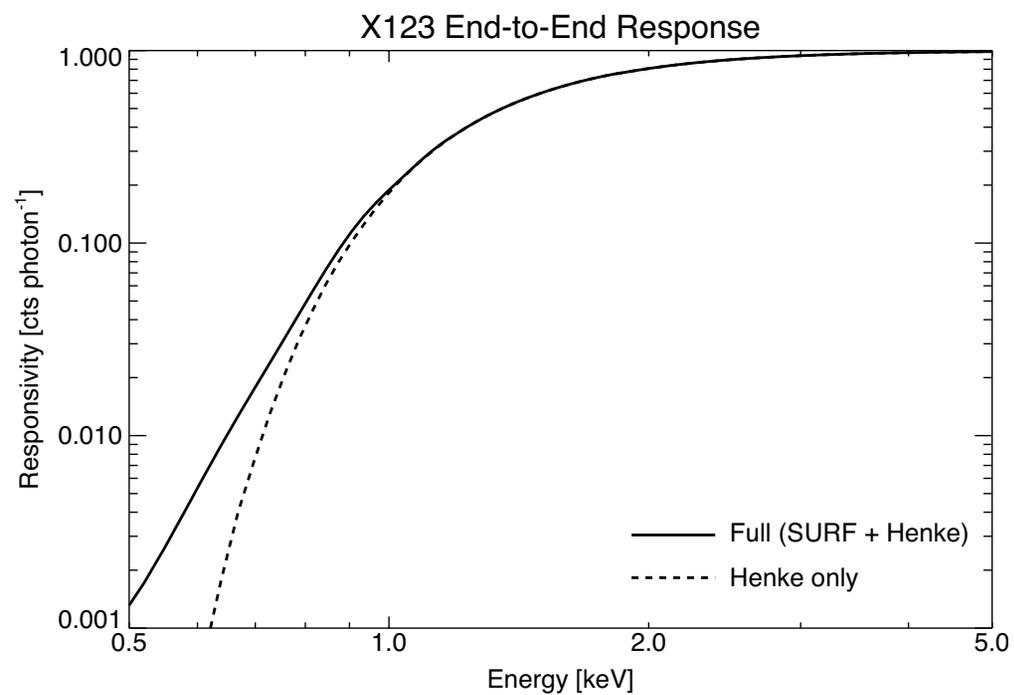

**Figure 1.** SURF calibration of X123. The instrument response (*right*) was derived in two pieces: above ~1.1 keV, by convolving the known SURF input spectrum with a Henke-model response (*dashed*) and fitting to the beam-normalized observed spectrum (*left, black*) from ~1.1 to 3.0 keV; below ~1.1 keV, by deconvolving the instrument resolution from the observations and directly dividing by the input spectrum. The model (*left, red*) resulting from the hybrid response fits the observations well at all energies.

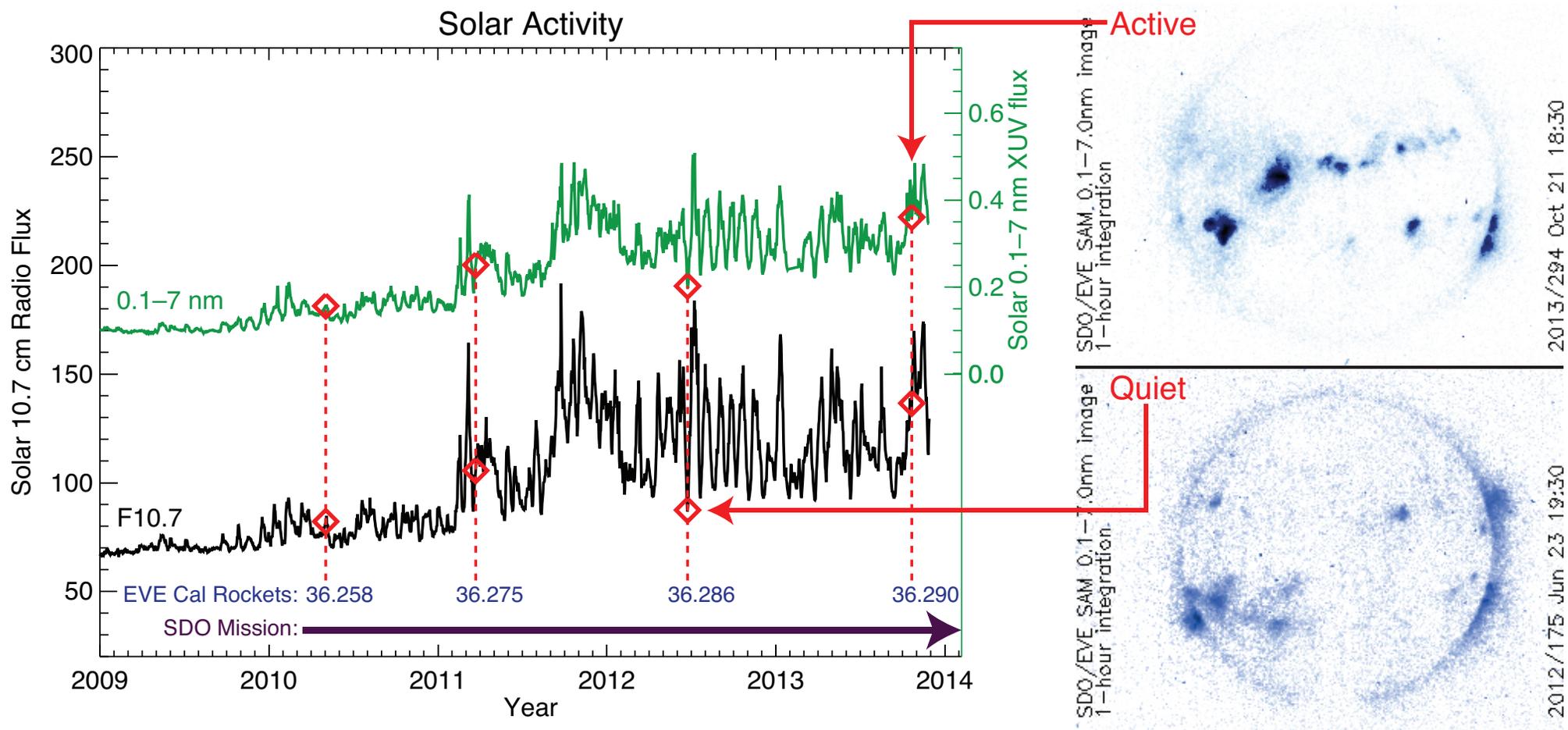

**Figure 2.** Activity levels during the two rocket flights. R20120623 (36.286) occurred during the minimum of a 27-day rotation, with only weak limb emission (*bottom right*), while R20131021 (36.290) occurred during an active period, with strong disk emission (*top right*).

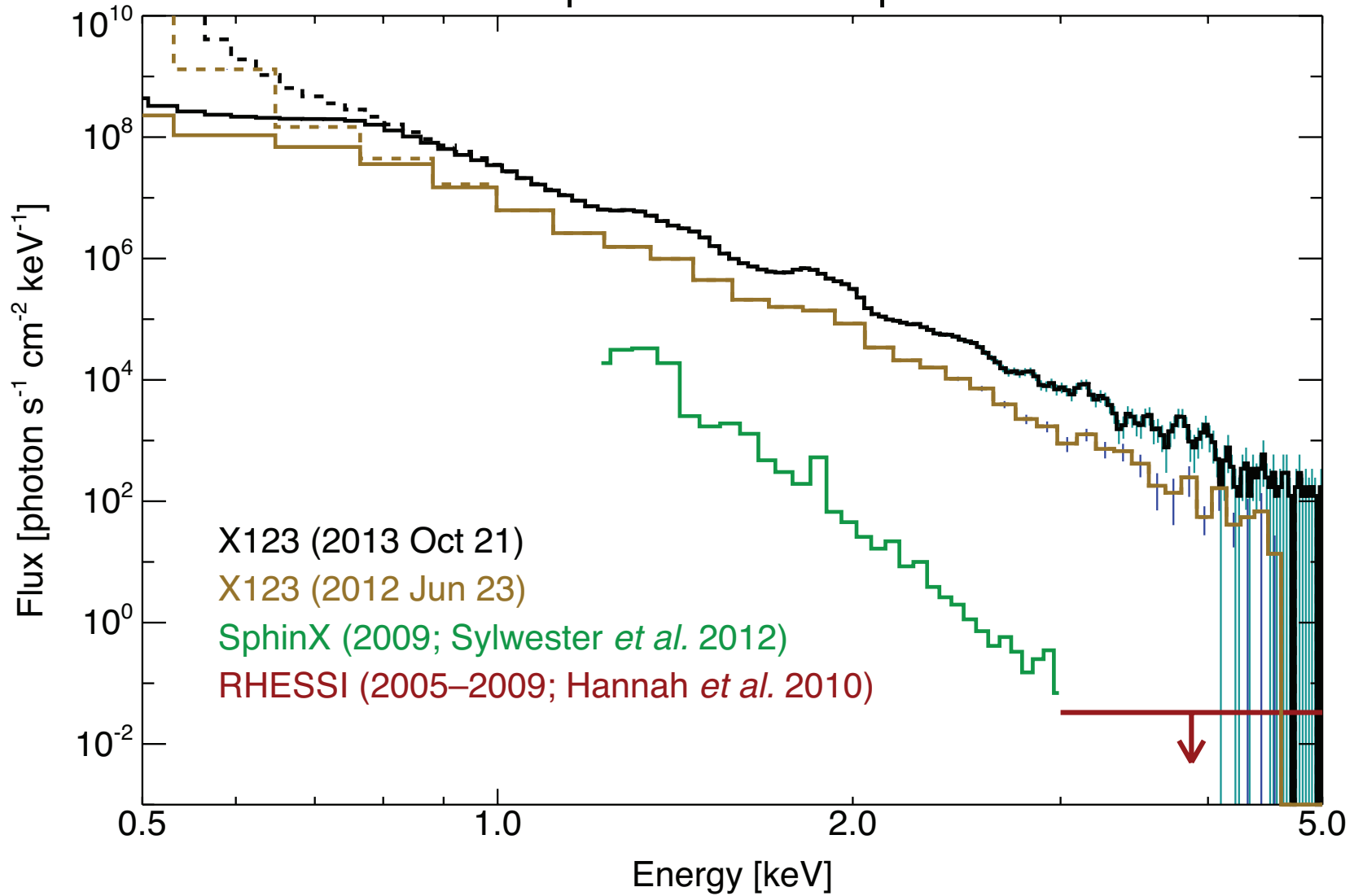

**Figure 3.** Spectral irradiance derived from the two rocket flights (*black, gold*) by dividing the observations by the hybrid instrument response (retaining the ~0.15 keV FWHM resolution); the Henke-only model (*dashed*) is an upper limit. Even the "quiet" observation of R20120623 is orders of magnitude higher than the 2009 deep-minimum observations by SphinX and the quiet Sun limits derived from *RHESSI*.

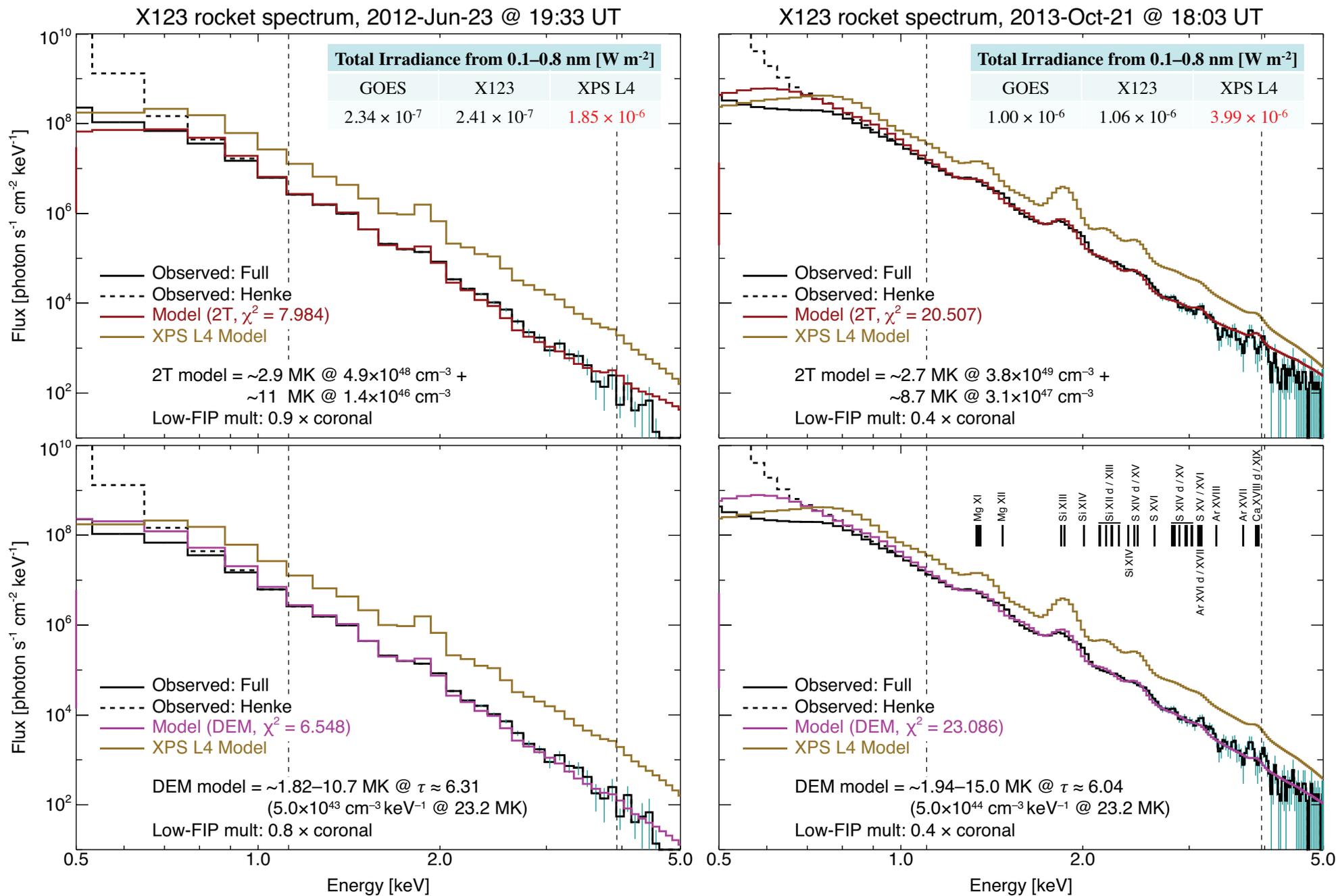

**Figure 4.** Spectral irradiances (*black*) and best-fit two-temperature (*top, red*) and DEM (*bottom, magenta*) models; R20131021 shows markedly stronger high-temperature emission and requires reduced low-FIP abundances. The high $\chi^2$ is acceptable given the simplistic models and large number of data points. XPS L4 (*gold*) overestimates the observations by ~7.7× and ~3.8×, respectively, while the integrated 0.1–0.8 nm irradiance derived from *GOES* XRS and from X123 agree to <10%.